\documentclass[journal]{IEEEtran}

\ifCLASSINFOpdf
\else
   \usepackage[dvips]{graphicx}
\fi
\usepackage{url}

\usepackage{amsmath,amsfonts,amssymb,bm}
\usepackage{siunitx}
\usepackage{hyperref}

\usepackage{graphicx}

\DeclareSIUnit{\nothing}{\relax}
\newcommand{\ours}{DIVE }
\newcommand{\Callhome}{CALLHOME{}}
\newcommand{\Mod}[1]{\ \mathrm{mod}\ #1}

\newcommand{\waveform}{x}
\newcommand{\numsamples}{L}
\newcommand{\numframes}{T}

\newcommand{\numspeakers}{N}
\newcommand{\numwindows}{W}
\newcommand{\dims}{D}
\newcommand{\speakerstack}{h}

\begin{document}

\title{DIVE: End-to-end Speech Diarization\\ via Iterative Speaker Embedding}

\author{Neil Zeghidour, Olivier Teboul and David Grangier
%\thanks{Submitted for review on DATE.}
\thanks{Google Research, Brain Team.}}

\maketitle

\begin{abstract}
We introduce DIVE, an end-to-end speaker diarization algorithm. Our neural algorithm presents the diarization task as an iterative process: it repeatedly builds a representation for each speaker before predicting the voice activity of each speaker conditioned on the extracted representations. This strategy intrinsically resolves the speaker ordering ambiguity without requiring the classical permutation invariant training loss. In contrast with prior work, our model does not rely on pretrained speaker representations and optimizes all parameters of the system with a multi-speaker voice activity loss. Importantly, our loss explicitly excludes unreliable speaker turn boundaries from training, which is adapted to the standard collar-based Diarization Error Rate (DER) evaluation. Overall, these contributions yield a system redefining the state-of-the-art on the standard \Callhome{} benchmark, with 6.7\% DER compared to 7.8\% for the best alternative.
\end{abstract}


\IEEEpeerreviewmaketitle

\section{Introduction}
\label{sec:intro}
Speech diarization is the task of annotating speaker turns in a conversation~\cite{DBLP:journals/taslp/TranterR06,DBLP:journals/taslp/MiroBEFFV12,DBLP:journals/corr/abs-2101-09624}. It is both a crucial step for downstream tasks such as automatic transcription of conversational speech, as well as a challenge as it requires handling long-term dependencies. Traditional systems typically split the problem in three sub-problems. First, a model is trained to extract short-term speaker embeddings. Such embeddings can be i-vectors derived from a Gaussian Mixture Model \cite{NingLTH06:spectral,ShumDDG13,ivector_diarization_1, ivector_diarization_2,DimitriadisF17,DBLP:conf/icassp/Kounades-Bastian17}, or embeddings produced by a neural network \cite{dvector_diarization_1, dvector_diarization_2, lin2019lstm,Kinoshita20,LiK0W21:discr_clustering}. Then, given a sequence to be diarized, a pre-trained speech activity detection algorithm~\cite{hughes:neural_vad,Thomas2015_vad} extracts active timesteps from the sequence and removes silences. Eventually, a clustering algorithm runs on top of these embeddings to assign each timestep to a speaker. Such composite systems have two main limitations. First, the speaker representations are not optimized for diarization, and may not extract relevant features for disambiguating speakers in e.g. presence of overlap. Moreover, most clustering algorithms being unsupervised, they cannot benefit from the fine-grained annotations of speaker turns in diarization datasets.

This motivated recent advances towards end-to-end diarization systems \cite{blstm_eend, von2019all}. In particular, \cite{blstm_eend, fujita2020end} propose to cast the diarization task as a multi-label classification problem. By training a model to predict whether each speaker is active at each timestep, a single model jointly performs speech activity detection (silence vs speech), speaker modelling and clustering. This framework has been used to train various architectures including LSTMs \cite{blstm_eend} and self-attention \cite{vaswani2017attention} models \cite{fujita_sa_eend_954}. Since diarization is a permutation-invariant problem (any permutation of the predicted speakers is valid), these models use Permutation-Invariant Training (PIT) \cite{yu17:pit,fujita2020end,xue2021:online} to avoid penalizing the model for choosing a particular speaker ordering. \cite{wavesplit} has shown that PIT can suffer from inconsistent assignments when applied to long sequences, and that it is preferable to explicitly learn long-term speaker representations. Moreover, fine-grained annotations can be unreliable around speaker turn boundaries, such that the standard is to remove the neighborhood of boundaries from evaluation \cite{lin2019lstm}. As a consequence, inconsistent supervision around boundaries
during training can adversely affect the final accuracy of the system.

In this work we introduce \ours (Diarization by Iterative Voice Embedding), an end-to-end neural diarization system. \ours combines three modules which are trained jointly: projection of the waveform to an embedding space, iterative selection of long-term speaker representations, and per-speaker per-timestep voice activity detection. The iterative speaker selection process addresses the problem of speaker order ambiguity and removes the need for training with PIT, similarly to attractor-based approaches~\cite{horiguchi_sa_eend_eda_807,fujita2020chainrule}. Moreover, we introduce collar-aware training, a modification to the standard multi-label classification loss which ignores errors in a defined radius around speaker turn boundaries to match the evaluation setting. \ours obtains a state-of-the-art Diarization Error Rate (DER) of $6.7\%$ on \Callhome \cite{callhome}. We also perform ablation studies that demonstrate the benefits of collar-aware training, and analyze the patterns of errors of our system.

\section{Method}
\label{sec:method}

\subsection{Setting and notations}
We consider a single channel recording $\waveform \in \mathbb{R}^{\numsamples}$ of $\numspeakers$, partially overlapping, speakers, with $\numsamples$ the length of the sequence. The goal of speech diarization is to produce per-speaker voice activity masks $y_i \in \{0,1\}^{T}$ for $i=1,\ldots,\numspeakers$, with $y_{i,t} = 1$ meaning that speaker $i$ is active at time $t$, and conversely. Typically, $ \numframes < \numsamples$ as the model does not produce voice activity masks at the sampling rate of the audio but rather at a lower sampling rate, e.g. every millisecond. \ours cascades three components. First, a {\it temporal encoder} projects the input waveform to a downsampled embedding space. Then, the {\it speaker selector} identifies one embedding (the speaker vector) that characterizes well each speaker, in an iterative fashion. Eventually, the {\it voice activity detector} consumes the embeddings produced by the temporal encoder as well as the selected speaker vectors and produces a binary voice activity mask for each speaker. We train these three modules jointly. In the following, we describe each component.

\begin{figure}
\centerline{\includegraphics[width=0.7\columnwidth]{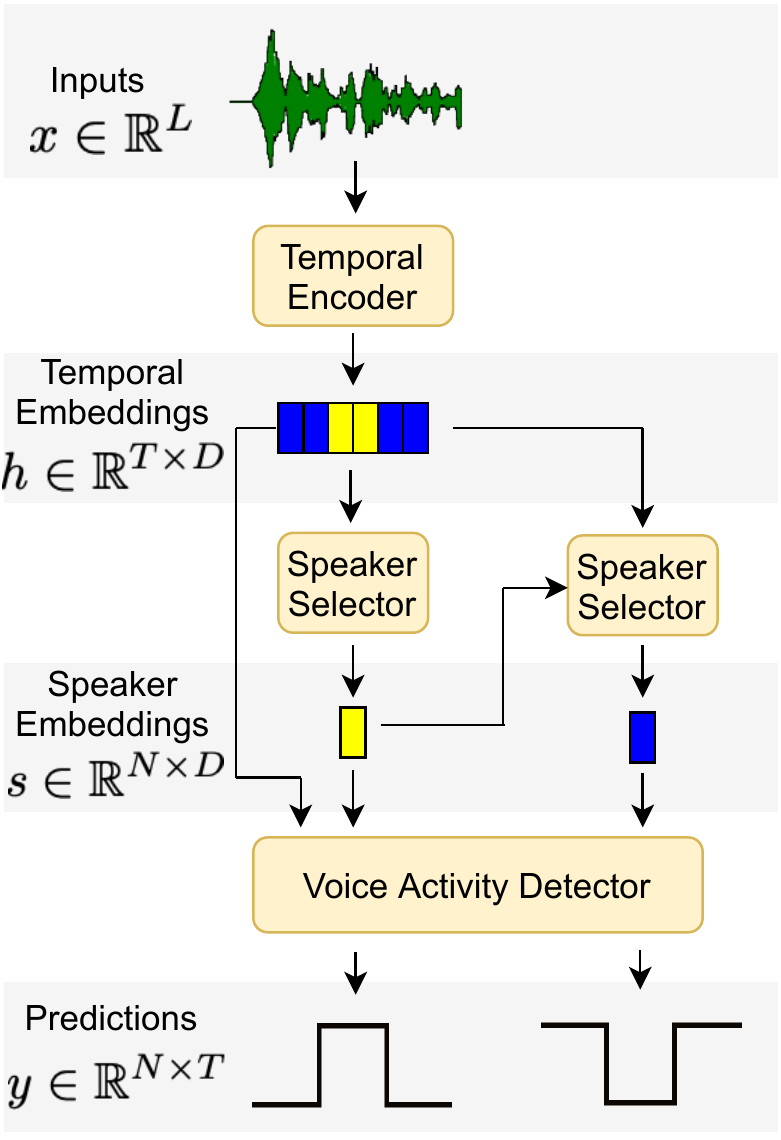}}
\caption{DIVE for 2 speaker diarization. The temporal encoder first extracts a local speaker representation. The speaker selector iteratively selects the representation of a novel speaker when a single speaker is active.
The voice activity module predicts speaker activity conditioned on 
the input signal and the selected representation.}
\end{figure}
\subsection{Temporal encoder}

The {\it temporal encoder} projects the input waveform $\waveform$ to an embedding space, while performing downsampling. Precisely, the temporal encoder $\speakerstack$ produces $\numframes$  latent vectors of dimension $\dims$, i.e. $h(x) \in \mathbb{R}^{\numframes \times \dims}$. We refer to these vectors as {\it temporal embeddings}. We use a temporal encoder similar to that of Wavesplit \cite{wavesplit}, which cascades residual blocks of dilated 1D-convolutions, with PReLU activations \cite{he15prelu} and Layer Normalization \cite{ba2016:layernorm}. Unlike Wavesplit \cite{wavesplit}, in which the temporal encoder maintains the original sampling rate of the signal, our temporal encoder performs downsampling by introducing 1D average pooling layers between residual blocks.
As the length of the audio sequence $\numsamples$ varies between examples, training on batches requires either truncating or padding sequences to a standard length. Given a batch of sequences, a typical training scheme is to randomly sample a fixed-length window from each sequence and to batch the resulting segments \cite{oord:wavenet, wavesplit}. As such windows are typically short (a few seconds), they are likely to only contain one to two speaker turns. This is not appropriate for training a diarization system that needs to model transitions between speaker turns and maintain long term consistency in speaker assignments. To address this issue, we instead sample $\numwindows$ fixed-length windows per sequence, pass them through the temporal encoder, and then concatenate them along the temporal axis. This allows for more diversity and more speaker turns inside a single training example. 
Section \ref{sec:multi_window} assesses the advantage of multi-window training.

\subsection{Iterative speaker selector}

The iterative speaker selector outputs a {\it speaker embedding} for each speaker detected in the signal. Iteratively, it takes as input the temporal embeddings $h$ and a representation of the previously selected speakers $\mu$ and outputs a representation of a single non-selected speaker $s$ along with a confidence score $c$. This process is repeated for two iterations in our two-speaker experiments but can be generalized to a variable number of iterations, stopping when the confidence drops below a given threshold. The representation of the previously selected speakers $\mu$ is defined recursively as the average of the embeddings of each previously selected speaker, starting with the zero vector for the first iteration. At each iteration $i$, this average embedding $\mu_i$ is mapped to a 4-by-\dims~matrix $g_{\mu}(\mu_i)$ with a fully-connected network. This matrix represents a 4-class linear classifier to map each temporal embedding $h_t$ to an event type $e_t$ among four possibilities: a single novel speaker is active, a single already selected speaker is active, overlapped speech, and silence, i.e.
$$
{\rm P}(e_t |h_t, \mu_i) = {\rm softmax}(g_{\mu}(\mu_i) g_{h}(h_t))
$$
where $g_h$ is a fully-connected network. At test time, the confidence of a selection iteration corresponds to the maximal confidence in the presence of a new speaker, i.e.
$$
c_i(t^\star_i) = \max_t c_i(t) = \max_t {\rm P}(e_t = {\rm novel~speaker} |h_t, \mu_i)
$$
and the speaker embedding corresponds to the temporal embedding where the maximal confidence is reached, i.e. $s_i = h_{t^\star_i}$. During training, we do not output $h_{t^\star_i}$ but instead a vector $h_t$ sampled uniformly from times with a novel speaker marked as active in the labels. This allows visiting a larger set of speaker representations. The learning process is supervised and the parameters of the model minimize the negative log likelihood of the 4-way classifier, 
$$
{\cal L}_{\rm selector}(h, \mu) = - \frac{1}{\numframes\numspeakers} \sum_{t=1}^{\numframes} \sum_{i=1}^{\numspeakers} 
\log {\rm P}(e_t |h_t, \mu_i).
$$

\subsection{Voice activity detector}
After selecting speaker embeddings, the last module of \ours predicts the voice activity of each speaker $y_i \in \{0,1\}^{T}$ for $i=1,\ldots,\numspeakers$. The voice activity detector contains two parallel fully-connected neural networks $f_h$ and $f_s$ with PReLU \cite{he15prelu} activations and Layer Normalization \cite{ba2016:layernorm}, except for the last layer which is a linear projection. To produce the voice activity $y_{i,t}$ of speaker $i$ at timestep $t$, $f_h$ and $f_s$ project the current temporal embedding $h_t \in \mathbb{R}^{\dims}$ and the speaker vectors $[s_i;\overline{s}] \in \mathbb{R}^{2\dims}$ respectively:
\begin{equation}
    \hat{y}_{i,t} = f_h(h_t)^\intercal f_s([s_i;\overline{s}]).
\end{equation}
Here,  $[s_i;\overline{s}]$ is the concatenation along the channel axis of $s_i$, the speaker vector of speaker $i$, and $\overline{s} = \frac{1}{\numspeakers} \sum_{j=1}^{N} s_j$ the mean of all speaker vectors. Intuitively, this means that when predicting the voice activity of a speaker at a given time, we use three pieces of information: the temporal embedding that represents the current speech content, a speaker embedding that represents the identity of the speaker of interest, and another embedding that represents all speakers. The latter allows the classifier to exploit contrasts between the current speaker of interest and other speakers in the sequence.

During training, we cast the problem of per-speaker, per-timestep voice activity detection as independent binary classification tasks and backpropagate the following loss:
\begin{equation}
\label{eq:standard_training}
{\cal L}_{\rm vad}(\hat{y}, y) = - \frac{1}{\numframes\numspeakers} \sum_{t=1}^{\numframes} \sum_{i=1}^{\numspeakers} \log(\sigma(\hat{y}_{i,t}(2y_{i,t}-1))).
\end{equation}
\subsection{Collar-aware training}
When evaluating a diarization system in terms of DER, it is common to apply a {\it collar}, which is a tolerance around speaker boundaries such that the metric does not penalize the model for small annotation errors. A typical value for such a tolerance is $250$ms on each side of a speaker turn boundary ($500$ms in total). Since we evaluate the model in these conditions, it would be beneficial to train it in a similar fashion i.e. to ignore errors within the collar tolerance. Thus, and as an additional contribution to the \ours architecture, we propose a training scheme for supervised diarization systems. During training, when computing the loss of the voice activity detector, we remove the loss of frames that fall inside a collar from the total loss and backpropagate the resulting masked loss:
\begin{equation}
\label{eq:collar_aware_training}
{\cal L}^{\rm collar}_{\rm vad}(\hat{y}, y) = - \frac{1}{\numframes\numspeakers} \sum_{\substack{t=1 \\ t\not\in B_r}}^{\numframes} \sum_{i=1}^{\numspeakers} \log(\sigma(\hat{y}_{i,t}(2y_{i,t}-1))),
\end{equation}
with $B_r$ the set of frames that lie within a radius $r$ around speaker turn boundaries. The effect of {\it collar-aware training} is illustrated in Figure \ref{fig:der_collar}. In Section \ref{sec:collar_ablation}, we show that training with the same collar as used for evaluation substantially improves the DER of the system.
The total loss minimized by \ours is therefore:
\begin{equation}
{\cal L}_{\rm total} = {\cal L}_{\rm selector} + {\cal L}^{\rm collar}_{\rm vad},
\end{equation}
and is used to train jointly the temporal encoder, iterative speaker selector and voice activity detector.

\section{Experiments}
\label{sec:exp}
We train our models on the "Fisher English Training Speech" Part 1 \cite{fisher_part_1} and Part 2 \cite{fisher_part_2}, two datasets of conversational telephone speech. Since they contain clean sequences, we simulate noisy situations by adding background noise from the ``noise'' part of MUSAN \cite{musan}. More precisely, when sampling a training speech sequence, we also sample a random background noise. We renormalize the energy of both the speech and noise sequences, sample a gain uniformly in $[-20,20]$ dB and apply it to the background noise before adding it to the speech sequence. We evaluate our models on the two-speaker evaluation of \Callhome~ \cite{callhome}, a multilingual conversational speech dataset. Following the standard of \cite{fujita_sa_eend_954, horiguchi_sa_eend_eda_807, horiguchi_sota_784_overlap} we report Diarization Error Rates averaged over the 148 test sequences. However, and unlike \cite{fujita_sa_eend_954}, we do not fine-tune our model on the 155 sequences of the ``adaptation'' set, but rather use it for hyperparameter selection. DERs are computed using the pyannote library \cite{pyannote.metrics}.
\subsection{Hyperparameters}
The temporal encoder first reduces the length $T$ of temporal embeddings with a 1D-Convolution with a kernel of size 16 and a stride of 8. It then cascades 4 blocks of 10 dilated convolution layers with kernel size 3 and stride 1. The dilation factor $\delta_l$ at layer $l$ follows the pattern of \cite{oord:wavenet, wavesplit}, i.e. $\delta_l = 2^{l \Mod{10}}$, which means that we reinitialize the dilation factor at the beginning of each block. Between the first two blocks, we perform average pooling with kernel size 3 and stride 2. Thus, the total downsampling factor of the model is 16 ($T=L/16$). All convolutional layers use 512 feature maps. The two branches $g_\mu$ and $g_h$ of the iterative speaker selector, as well as those ($f_h$ and $f_s$) of the voice activity detector have two hidden layers with 512 feature maps. We train our model with Adam \cite{kingma14:adam} and a batch size of 512, using an initial learning rate of 0.0003, decayed by a factor of 0.7 every 50000 batches. We use multi-window training with 6 windows of length 32000 samples each.

\begin{table}[t]
    \centering
    \caption{Diarization Error Rate (DER) in $\%$ on the test set of \Callhome. All models are evaluated with a 250ms collar. "NO OVERLAP" means that the evaluation excludes overlapped speech.}
    %\vspace{0.2cm}
    \label{tab:der_\Callhome}
    \begin{tabular}{l|c|c}
            \hline\hline
            \multicolumn{1}{c|}{Model} & OVERLAP & NO OVERLAP\\
            \hline\hline
            UIS-RNN V1 \cite{zhang_fully_supervised} & -- & 10.6 \\
            UIS-RNN V2 \cite{zhang_fully_supervised} & -- & 9.6 \\
            UIS-RNN V3 \cite{zhang_fully_supervised} & -- & 7.6 \\
            x-vector + LSTM \cite{lin2019lstm} & -- & 6.6 \\
            BLSTM-EEND \cite{blstm_eend}  & 23.1 & -- \\
            SA-EEND \cite{fujita_sa_eend_954}  & 9.5 & -- \\
            SA-EEND-EDA \cite{horiguchi_sa_eend_eda_807}  & 8.1 & -- \\
            SA-EEND-EDA + Frame Selection \cite{horiguchi_sota_784_overlap}  & 7.8 & -- \\
            \hline
            \ours   &\bf{6.7} & \bf{5.9} \\
            \hline                
    \end{tabular}
\end{table}

\subsection{\Callhome}
Table \ref{tab:der_\Callhome} reports the DER on the test set of \Callhome. The UIS-RNN \cite{zhang_fully_supervised} is an hybrid system training an RNN on top of pre-trained speaker embeddings, with the V3 being trained on a proprietary dataset with \SI{138}{\kilo\nothing} speakers. Similarly, \cite{lin2019lstm} trains an LSTM to model the similarity between pre-trained speaker embeddings and performs diarization. Both models are evaluated without considering overlapped speech, and the latter uses oracle speech activity labels (removing silences). Table \ref{tab:der_\Callhome} shows that \ours outperforms both systems in this condition, reaching $5.9\%$ DER, even though \ours is trained in an end-to-end fashion, without any speaker label and without oracle speech activity annotations. BLSTM-EEND \cite{blstm_eend} trains a bidirectional LSTM for per-speaker per-timestep voice activity detection, with an additional speaker clustering loss, similar in spirit to \ours, with a DER of $23.1\%$. SA-EEND \cite{fujita_sa_eend_954} replaces the LSTM by self-attention \cite{vaswani2017attention} and removes the deep clustering loss, with the best variation reaching $7.8\%$ thanks to vast training data and fine-tuning on the \Callhome~ validation set. \ours outperforms these models, with a $6.7\%$ DER, and despite not being fine-tuned on \Callhome. In Table \ref{tab:der_\Callhome}, the results for \ours are obtained with a 11-frame median filtering on top of the predictions of the model, as suggested in \cite{blstm_eend}. This avoids predicting non-existing, extremely short segments. Without this median filtering, the DER of \ours goes up from $6.7\%$ to $6.8\%$, which shows that the model's predictions are already reliable.

\begin{figure}
    \centering
    \includegraphics[width=0.4 \textwidth]{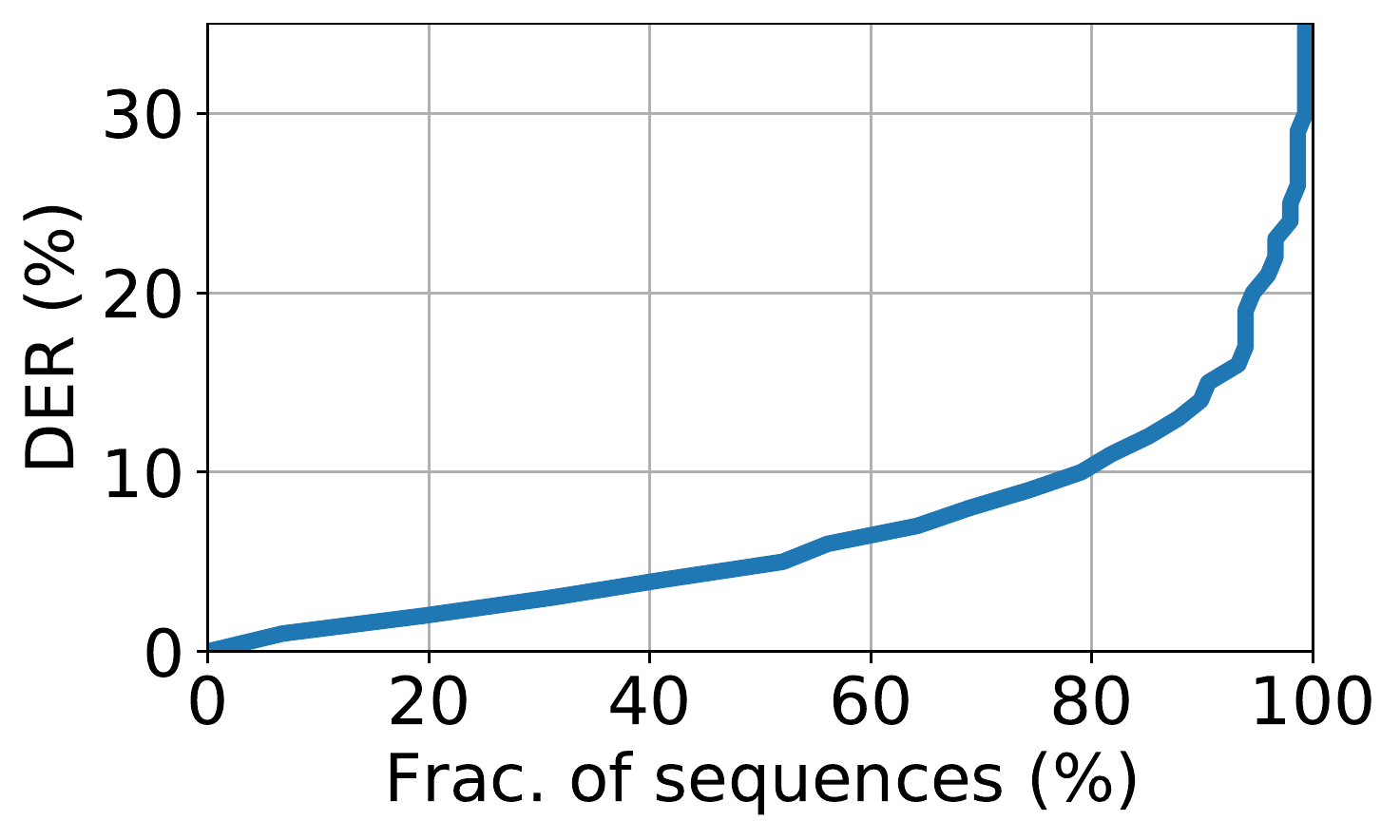}
    \caption{Cumulative distribution of the Diarization Error Rate (DER) in \% on \Callhome, with the standard 250ms collar.}
    \label{fig:der_cumul}
\end{figure}

Table \ref{tab:contingency_\Callhome} analyzes error types. We observe few confusion errors where a speaker is mistaken for another (1.5\%); most errors concentrate on mistaking single speaker activity for overlapped speech (7.4\%), mistaking overlap for single speaker activity (3.3\%) and mistaking silence for speaker activity (5.6\%). Figure~\ref{fig:der_cumul} plots the cumulative distribution of DER and shows that the median DER is below 5 while the average is higher due to a minority (5\%) with DER over 20.
\begin{table}[t]
    \centering
    \caption{Labels vs Predictions Contingency (\%) for frame-wise diarization on \Callhome.}
    %\vspace{0.2cm}
    \label{tab:contingency_\Callhome}
    \begin{tabular}{l l | r| r | r| r}
    \hline\hline
        & & \multicolumn{4}{c}{Labels}\\
        & & Spkr. 1 & Spkr. 2 & Overlap & Silence\\\hline \hline
Prediction & Spkr. 1 & {\bf 49.6} & 0.9 & 1.8 & 3.5\\ 
         & Spkr. 2 & 0.6 & {\bf 18.8} & 1.5 & 2.1\\ 
         & Overlap & 4.1 & 3.3 & {\bf 8.4} & 0.9\\ 
         & Silence & 0.7 & 0.4 & 0.0 & {\bf 3.3}\\  \hline
\multicolumn{2}{l|}{Class prior} & 55.1 & 23.3 & 11.8 & 9.8\\          \hline
    \end{tabular}
\end{table}

\subsection{Impact of collar-aware training}
\label{sec:collar_ablation}
\begin{figure}
    \centering
    \includegraphics[width=0.48 \textwidth]{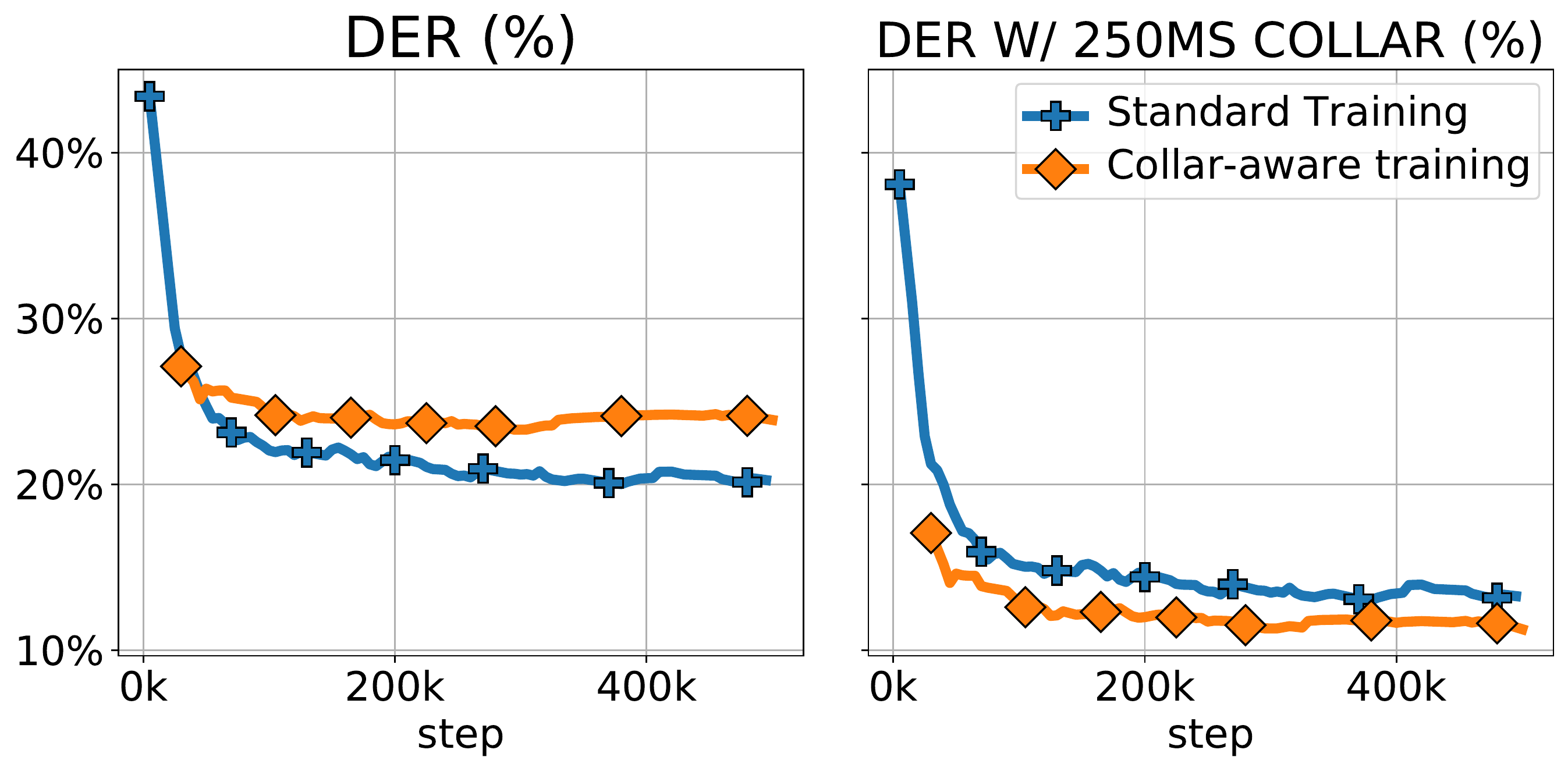}
    \caption{Diarization Error Rate (DER) in \% with and without collar-aware training, on the validation set of \Callhome. On the left is the raw DER, that penalizes every error. On the right, the DER with the standard 250ms collar.}
    \label{fig:der_collar}
\end{figure}

Figure \ref{fig:der_collar} shows the impact of collar-aware training. When using the standard loss function of Equation \ref{eq:standard_training}, the raw DER decreases steadily. On the other hand, when using the collar-aware loss defined in Equation \ref{eq:collar_aware_training}, the raw DER plateaus early in training, but its DER with a 250ms collar converges faster and to a better score than its standard counterpart. This shows that when the target evaluation metric uses a collar, it is beneficial to integrate this tolerance into the training loss.

\subsection{Impact of training on multiple windows}
\label{sec:multi_window}

\begin{figure}
    \centering
    \includegraphics[width=0.40 \textwidth]{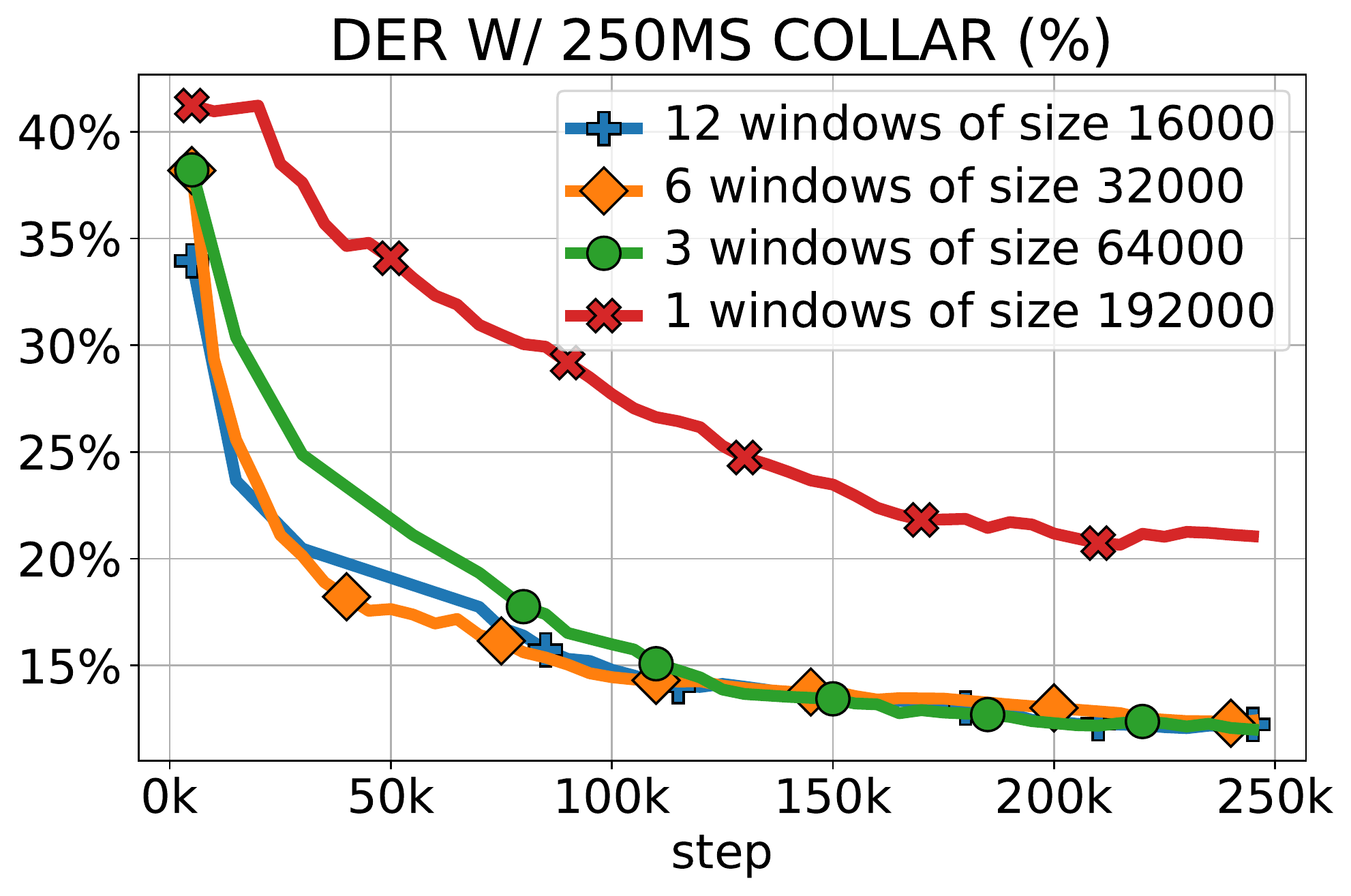}
    \caption{Diarization Error Rate (DER) in \% when varying the number and size of windows for a constant total of 192000 samples.}
    \label{fig:multi_window}
\end{figure}

Diarization requires speaker representations that are reliable throughout long sequences of speech, e.g. several minutes. However, training a neural network over long speech sequences is slow since it prevents from training with a large batch size due to memory constraints. To solve this problem, our training process samples multiple short windows from the same sequence: this allows \ours to observe the same speakers over snippets far apart in time while keeping memory usage low. Figure~\ref{fig:multi_window} illustrates the benefit of multi-window training: for a fixed budget of $192000$ samples per training example, splitting it into several windows performs much better than using a single, contiguous window.

\section{Conclusions}
\label{sec:ccl}

This paper introduces \ours, an end-to-end model for speaker diarization. \ours decomposes the task into three stages: convolutional temporal encoding, iterative 
speaker selection and speaker-conditioned voice activity prediction. The iterative speaker 
selector repeatedly processes the whole sequence to select a representation of a 
speaker not selected during the previous iterations. The extracted representations condition voice activity prediction. This formulation resolves the ambiguity in speaker order and offers a generic formulation regardless of the number of speakers per sequence.
The model does not rely on pretrained components and all parameters are trained to 
optimize the voice activity likelihood with a novel collar-aware loss function. This  loss does not rely on supervision from unreliable speaker turn boundaries, and matches standard collar-aware evaluations. \ours establishes a new state-of-the-art on the standard \Callhome{} benchmark, with 6.7\% DER compared to
7.8\% for the best alternative.
In the future, we aim to address experimental settings with variable number of
speakers and noisier acoustic conditions~\cite{dihard3,chime6}.

\clearpage
\newpage

\vspace{-2mm}
\bibliography{main}
\bibliographystyle{IEEEtran}

\end{document}